\documentclass[preprint,prb,showpacs,preprintnumbers,amssymb]{revtex4}
\usepackage{latexsym}
\usepackage[dvips]{graphicx}
\usepackage{bm}
\usepackage{epsf}
\usepackage{epsfig}
\usepackage{dcolumn}
\begin{document}
\title{One-dimensional Fe surface states confined by
self-assembled carbon chains at the Fe(001) surface}
\author{G. Trimarchi}
\affiliation{INFM DEMOCRITOS National Simulation Center and
The Abdus Salam International Center for Theoretical Physics,
Strada Costiera 11, 34014 Trieste, Italy}
\author{N. Binggeli}
\affiliation{INFM DEMOCRITOS National Simulation Center and
The Abdus Salam International Center for Theoretical Physics,
Strada Costiera 11, 34014 Trieste, Italy}
\date{\today}
\begin{abstract}
A c($3\sqrt{2} \times \sqrt{2}$) reconstructed structure of the Fe($001$)
surface with self-assembled C zigzag chains has been recently observed
experimentally.
Using {\em ab initio} density-functional calculations, we address the 
effect of the C zigzag chains on the electronic structure of this surface.
We find that the formation of the C chains produces one-dimensional surface 
states localized along the zigzag chains. 
These states are spin-polarized and derive from preexisting two-dimensional 
Fe $d_{3z^{2}-r^{2}}$-like surface states of the clean Fe(001) surface.
The simulation of the STM image allows us to assign the chain-like 
structure, observed in STM experiments, to the one-dimensional 
Fe surface states laterally confined within the C zigzag stripes.
\end{abstract}

\maketitle
The confinement of metal surface states using lateral nano or 
atomic-scale structures has attracted much interest in recent years. 
Well-known examples are ``quantum corrals'', where free-electron-like 
standing waves of a metal surface are confined within corral-like 
rings composed of ad-atoms.\cite{Eigler93} 
Other striking examples   
include confined surface states at vicinal or decorated stepped 
surfaces \cite{Avouris94} and one-dimensional surface states at 
domain boundaries in two-dimensional alloys.\cite{Biedermann96}  
The study of metal surface states and their manipulation via lateral 
structures is not only of fundamental interest, i.e., to investigate 
quantum interferences and/or low-dimensional systems, but also of 
practical importance, since surface states are known to influence the 
adsorption characteristics of surfaces and, in the case of 
transition-metal surfaces, also the surface magnetism.\cite{Memmel98} 

Self-assembling has been the focus of extensive investigations lately, 
as a promising tool to build ordered nano or atomic-scale structures, 
without the direct manipulation of atoms or molecules. 
This approach has been used, e.g., to grow semiconductor quantum 
dots\cite{Vdovin00} and semiconductor and metal quantum 
wires\cite{Fuster04} on semiconductor surfaces.
In general, in self-assembled systems, the confinement 
concerns the electronic states of the self-organized species. 
The possibility, however, to manipulate surface states of the 
substrate, and in particular to laterally confine them within regular 
patterns via self-organized structures, is also of special interest 
as a potential approach to engineer metal surface properties. 

Very recently, an interesting case of self-organized chains of 
non-magnetic impurities on a magnetic metal was reported. 
A Borges-garden-like arrangement of C stripes on Fe(001), 
corresponding to a C coverage of 2/3 monolayer (ML), was induced  
via surface segregation of bulk C impurities.\cite{Fujii03,Pana05} 
Scanning tunneling microscopy (STM) indicated the presence of 
self-assembled C zigzag chains forming a regular pattern over 
a nanometer lateral scale, corresponding to a 
$c(3\sqrt{2}\times \sqrt{2})$ reconstructed structure. 

In the present work, using state-of-the-art {\em ab initio} calculations, 
we examine this $c(3\sqrt{2} \times \sqrt{2})$ reconstructed structure,  
and show that the formation of the C chains gives rise to one-dimensional 
Fe surface states close to the Fermi energy. 
These states are spin-polarized and laterally confined within the C 
zigzag stripes. We identify the origin of their confinement, and show 
that such surface states are responsible for the chain-like 
structures observed in STM experiments. 
We also predict that the one-dimensional Fe states give rise to a 
significant feature in the local density of states, which should be 
measurable by scanning tunneling spectroscopy.
The surface states we predict exhibit some similarities with the 
localized Fe states found in Ref.~\onlinecite{Biedermann96} at 
Fe-rich domain boundaries on the Si/Fe(001) surface alloy with  
$\sim35$\% Si. 
In the latter alloy system, however, the observed chains of Fe 
surface states are disordered, and originate from  
a different local atomic geometry, i.e., from Si occupying 
substitutional sites within the Fe(001) surface layer, whereas  
the C atoms segregate at hollow sites and produce self-organized  
stripes. 
\begin{figure}
  \begin{center}
    \epsfig{file=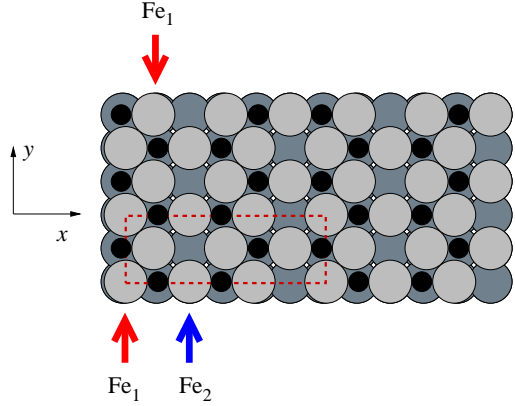, scale = 0.45}
    \caption[C/Fe(001) geometry]{\label{fig:geo}
      (Color online) Top view of the C/Fe(001) c($3\sqrt{2} \times \sqrt{2}$) surface. 
      The dashed box indicates the c($3\sqrt{2} \times \sqrt{2}$) surface 
      unit cell. Small (large) disks corresponds to C (Fe) atoms; light (dark)
      gray disks indicate top (second layer) Fe atoms.  The carbon atoms sit 
      on hollow sites and lie $\sim0.2$ \AA~above the outermost Fe atoms.
      Fe$_{1(2)}$ indicates Fe surface atoms enclosed within (in between) the
      C stripes.}
  \end{center}
\end{figure}

Our first-principle calculations are performed within density functional 
theory, using the spin-polarized Perdew-Burke-Ernzerhof\cite{pbe:theo}   
exchange-correlation functional and Vanderbilt\cite{vander:theo} 
ultra-soft pseudopotentials with a plane-wane basis set.
The kinetic-energy cutoffs for the wave functions and electronic
density are set to $27$ Ry and $270$ Ry, respectively. The sampling of
the Brillouin Zone (BZ) is done with a Monkhorst-Pack (MP) grid, and we use 
a Gaussian smearing of the electronic levels with a full-width at half maximum 
of $0.02$ Ry to determine the Fermi energy. The theoretical lattice constant 
$a$ of bulk iron is $2.86$ \AA\@ (compared to the experimental value of 
$2.87$ \AA). 
To model the C/Fe($001$) c($3\sqrt{2} \times \sqrt{2}$) surface, we use a 
symmetric slab of 9 Fe atomic layers with 4 C atoms per  
c($3\sqrt{2} \times \sqrt{2}$) surface unit.
The thickness of the vacuum regions separating the periodic images of the slab 
corresponds to $9$ Fe($001$) atomic layers (i.e., $\sim 12.85$ \AA)
and we use a (2,6,1) MP grid for the BZ integrations.
In Fig.~\ref{fig:geo}, we display the equilibrium atomic structure of the C/Fe(001)
with 2/3 ML of C. The zigzag-chain structure is the lowest-energy 
configuration we obtain for this coverage and reconstruction. 
The details of the structural investigation will be presented 
elsewhere;\cite{Pana05} it is only important here to note that 
the C atoms of the zigzag chains occupy hollow sites and display  
a strong relaxation towards the Fe surface layer, i.e, their 
equilibrium position is only $0.2$~\AA\@ higher than that of the 
outermost Fe atoms.

\begin{figure}
  \begin{center}
    \epsfig{file=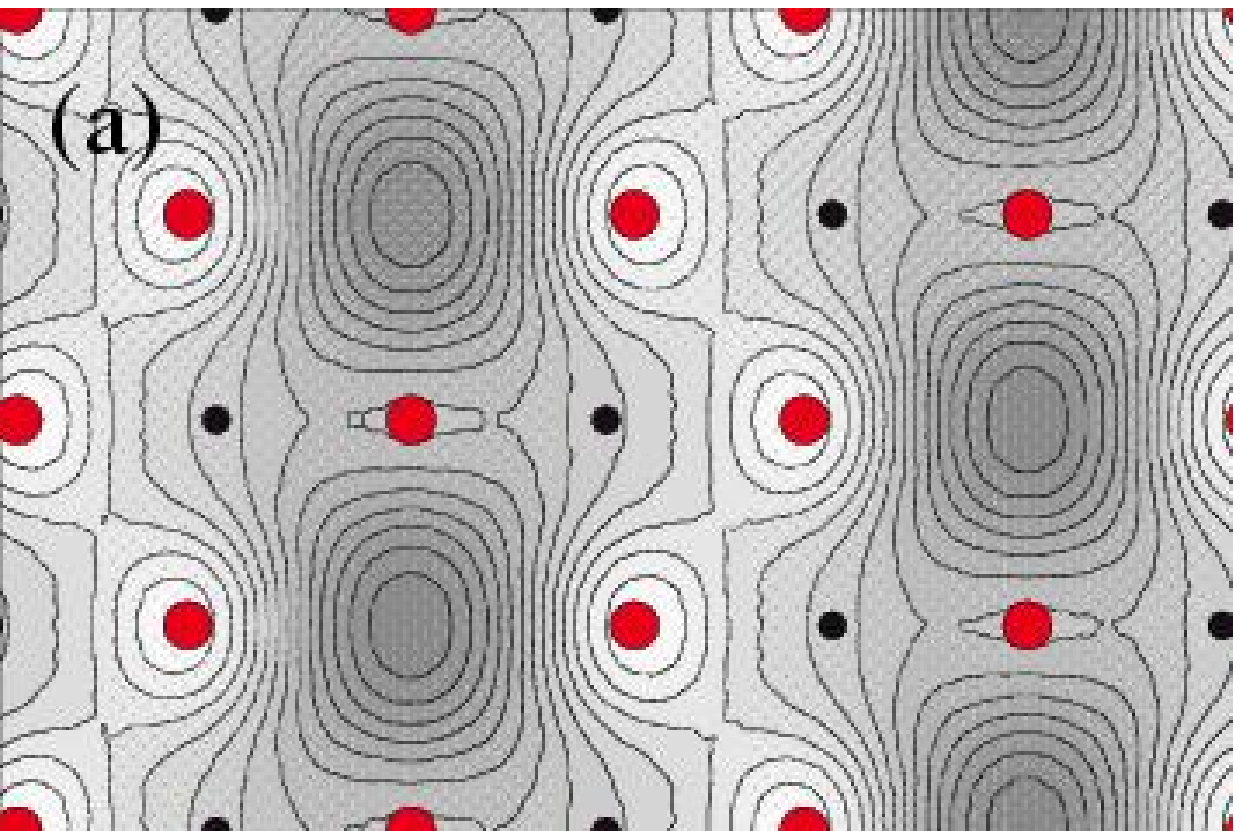,scale = 0.45}\vspace*{0.3cm}

    \epsfig{file=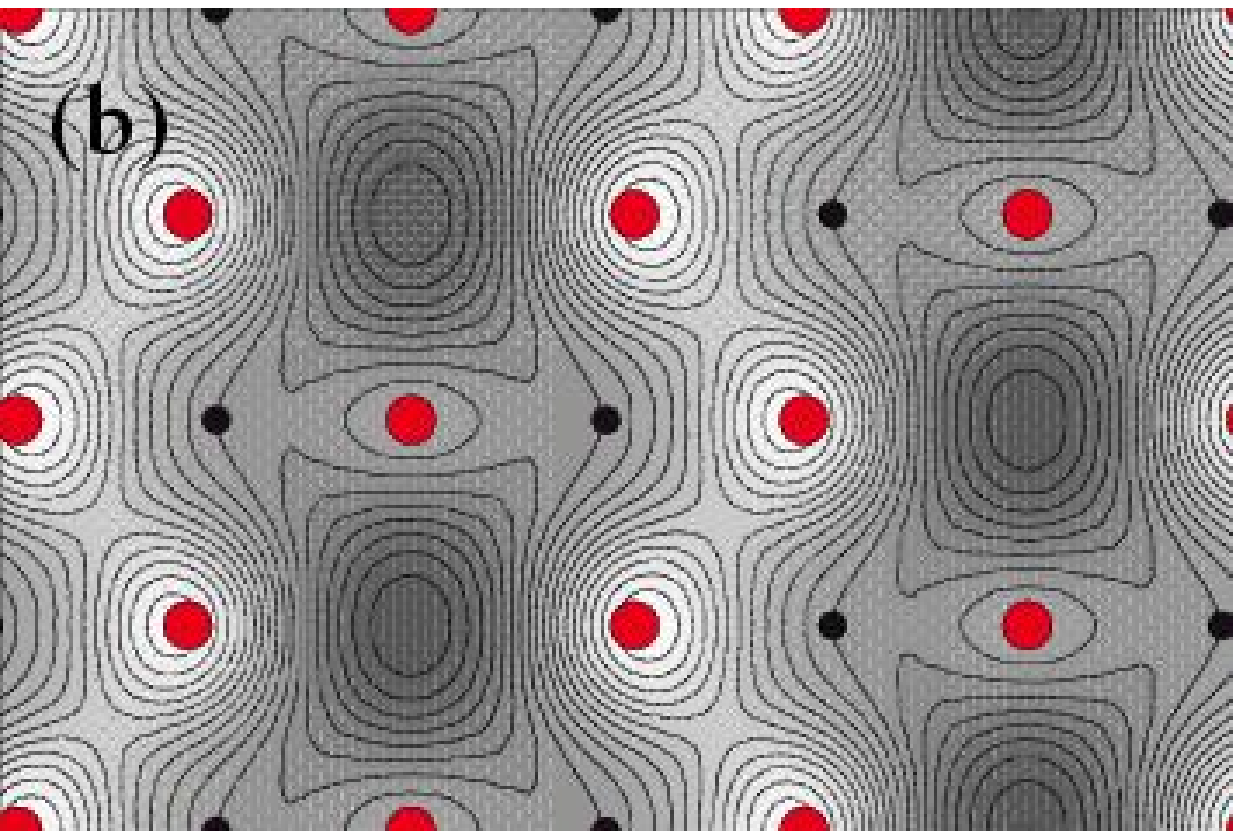,scale = 0.45}\vspace*{0.3cm}

    \epsfig{file=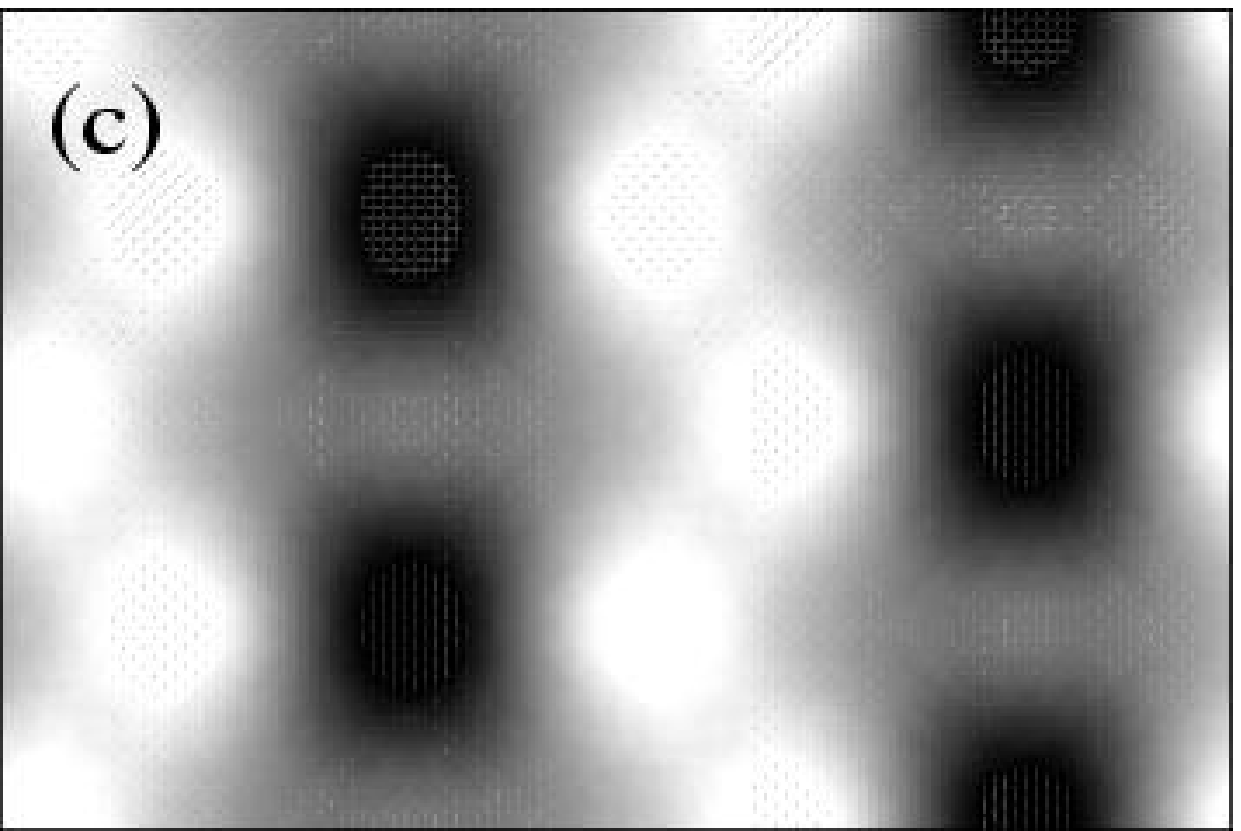,scale = 0.45}

    \caption[MnOMn angles]{\label{fig:stm} (Color online) 
        Contour plots in a $(001)$ plane, at a distance
 	d$\approx 2.6$ \AA\@ from the carbon surface layer, of the
	local density of states integrated over the energy window $[E_{F},E_{F}+\Delta 
        \varepsilon]$,
	where E$_{F}$ is the Fermi energy and $\Delta \varepsilon=$ 0.4 (a) and 0.9 (b) eV.
	The positions of the Fe atoms (large disks) and C atoms (small disks) are 
        also indicated.
	The spacing between the contours in (a) and (b) is respectively  
        $0.78 \times 10^{-6}$ e/bohr$^{3}$ and $3.2 \times 10^{-6}$ e/bohr$^{3}$, with the 
        contour of lowest density corresponding respectively to $5.9 \times 10^{-6}$ and 
        $4.9 \times 10^{-5}$ e/bohr$^{3}$. Panel (c) shows a gray scale plot of 
        the integrated local density of states obtained for $\Delta \varepsilon$ = 0.4 eV in 
        a ($001$) plane at d$\approx 3.0$ \AA: light/dark gray corresponds to large/small 
        values of the integrated local density of states.}
  \end{center}
\end{figure}

In connection with the interpretation of the STM experiments, 
we have examined the spatial behavior in the vacuum, near the 
surface, of the local density of empty electronic states 
integrated between the Fermi energy, $E_{\text{F}}$, and 
$E_{\text{F}} + \Delta \varepsilon$, with $\Delta \varepsilon$ 
taken as large as 1~eV and as low as 0.2~eV. 
In Fig.~\ref{fig:stm}a (\ref{fig:stm}b), we display the 
integrated local density of states, 
$I(\Delta \varepsilon, \bm{r})$, obtained for $\Delta \varepsilon 
= + 0.4$~eV ($+ 0.9$~eV), in a (001) plane located in the 
vacuum at a distance $d \approx 2.6$~\AA\@ from the outermost 
C layer. Within the Tersoff-Hamann approximation, 
$I(\Delta \varepsilon, d, \bm{r}_{\parallel})$ is proportional to 
the tunneling current corresponding to a bias  $\Delta \varepsilon$, 
tip distance $d$, and lateral position 
$\bm{r}_{\parallel}$.\cite{Tersoff83}  
Inspection of Fig.~\ref{fig:stm} indicates that the integrated local density 
of states has its maxima positioned on top of the Fe surface 
atoms enclosed within the C stripes (Fe$_1$ atoms in Fig.~\ref{fig:geo}). 
This gives rise to a zigzag chain structure in the simulated 
STM image at positive bias (see Fig.~\ref{fig:stm}c), in good agreement 
with the experimental STM images,\cite{Fujii03,Pana05} even though,  
counter-intuitively, bright spots do not correspond to the C atoms 
of the zigzag chains, but to the Fe atoms enclosed within their  
stripes. 
We note that for all  $\Delta \varepsilon$ examined, from 0.2 to 1~eV, 
and for all distances $d$ considered, from 2.6 to 3.2~\AA, the 
integrated local density of states exhibits a qualitatively similar 
behavior, i.e., a zigzag chain arrangement of maxima (bright spots) 
located on the Fe$_1$ atoms. 

\begin{figure}
  \begin{center}
    \epsfig{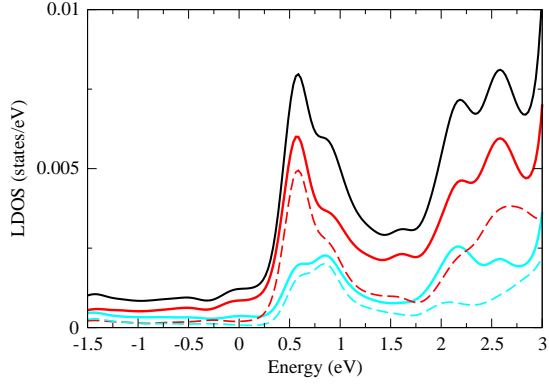}
    \caption[DOS]{\label{fig:ldos} 
      (Color online) Local Density of States integrated over the vacuum space 
      between a distance d$=3.2$~\AA, measured from the outermost C layer, 
      and the middle of the vacuum slab: d$\approx6.7$~\AA\@ (solid black line). 
      The red/blue (dark/light gray) solid line shows the contribution from the 
      areas enclosed within/between the C stripes, and the red/blue 
      (dark/light gray) dashed line the corresponding contribution from the 
      minority-spin states. The zero of energy 
      corresponds to the Fermi level.  The densities of states have been 
      convoluted with a Gaussian of standard deviation $0.1$~eV.}
  \end{center}
\end{figure}

To better understand this behavior, we have investigated the energy 
distribution and the $k$-space dispersion of the tunneling states. 
In Fig.~\ref{fig:ldos}, we display the local density of states (LDOS) integrated 
in the vacuum region between $d = 3.2$~\AA\@ and the middle of the 
vacuum slab ($d \approx 6.7$~\AA). The separate contributions from areas 
enclosed within and in-between the carbon stripes (with area ratio 2:1) are also 
shown, as well as the specific contributions from the minority-spin states 
enclosed within these regions.
The minority-spin states are responsible for a broad main feature,  
located at $\sim0.6$ eV above $E_{\text{F}}$, and a smaller feature, at 
$\sim0.8$~eV. The former (latter) is related mainly to states enclosed 
within (in-between) the C stripes. We note that the majority-spin 
states do not give rise to any significant feature up to $\sim2.0$ eV above 
$E_{\text{F}}$.

\begin{figure}
  \begin{center}
    \epsfig{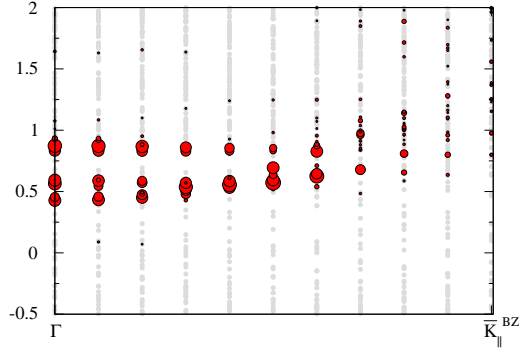}
    \caption[BANDS]{\label{fig:band} (Color online) Calculated surface band structure 
      of the minority-spin tunneling states with largest integrated probability
      density in the vacuum space beyond d$\approx 2.3$ \AA.
      The dispersion is shown along the $\bar{\Gamma} - \bar{K}_{\|}^{BZ}$ line 
      parallel to the C chains in the Brillouin zone of the c($3\sqrt{2} \times 
      \sqrt{2}$)surface; $\bar{K}_{\|}^{BZ}$ stands for 
      the zone-edge $k$ point $(\frac{\pi}{2a},\frac{\pi}{2a})$.
      The eigenvalues are indicated with gray points, while the filled  red 
      (dark gray) 
      circles indicate states that have a non-negligible charge density beyond 
      d$\approx2.3$ \AA. The radii of the circles are proportional to the
      integrated charge density of the corresponding states between $\sim2.3$ \AA\@ 
      and $\sim6.7$ \AA. The zero of energy corresponds to the Fermi level. }
  \end{center}
\end{figure}

\begin{figure*}
  \begin{center}
    \epsfig{file=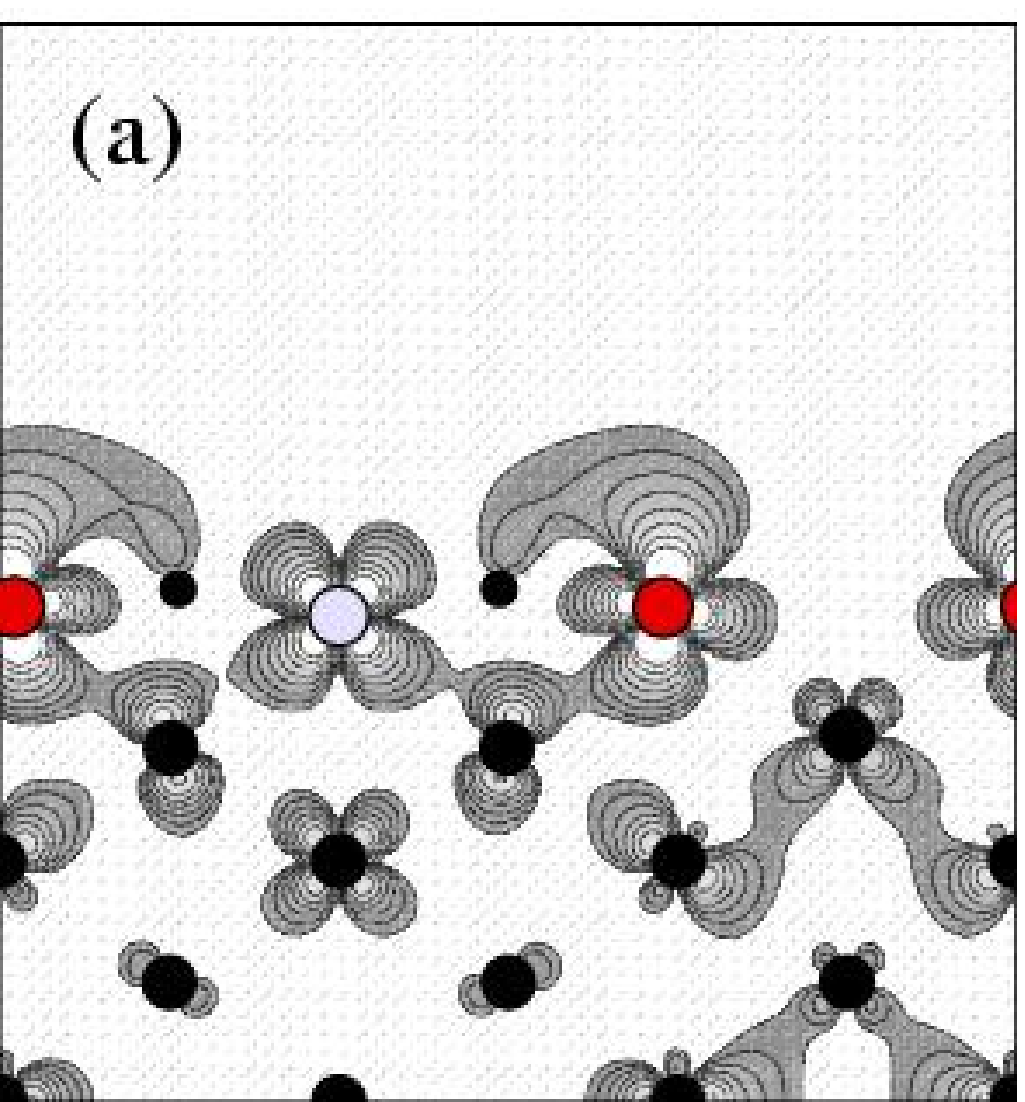, scale = 0.40}\hspace*{0.25cm}
    \epsfig{file=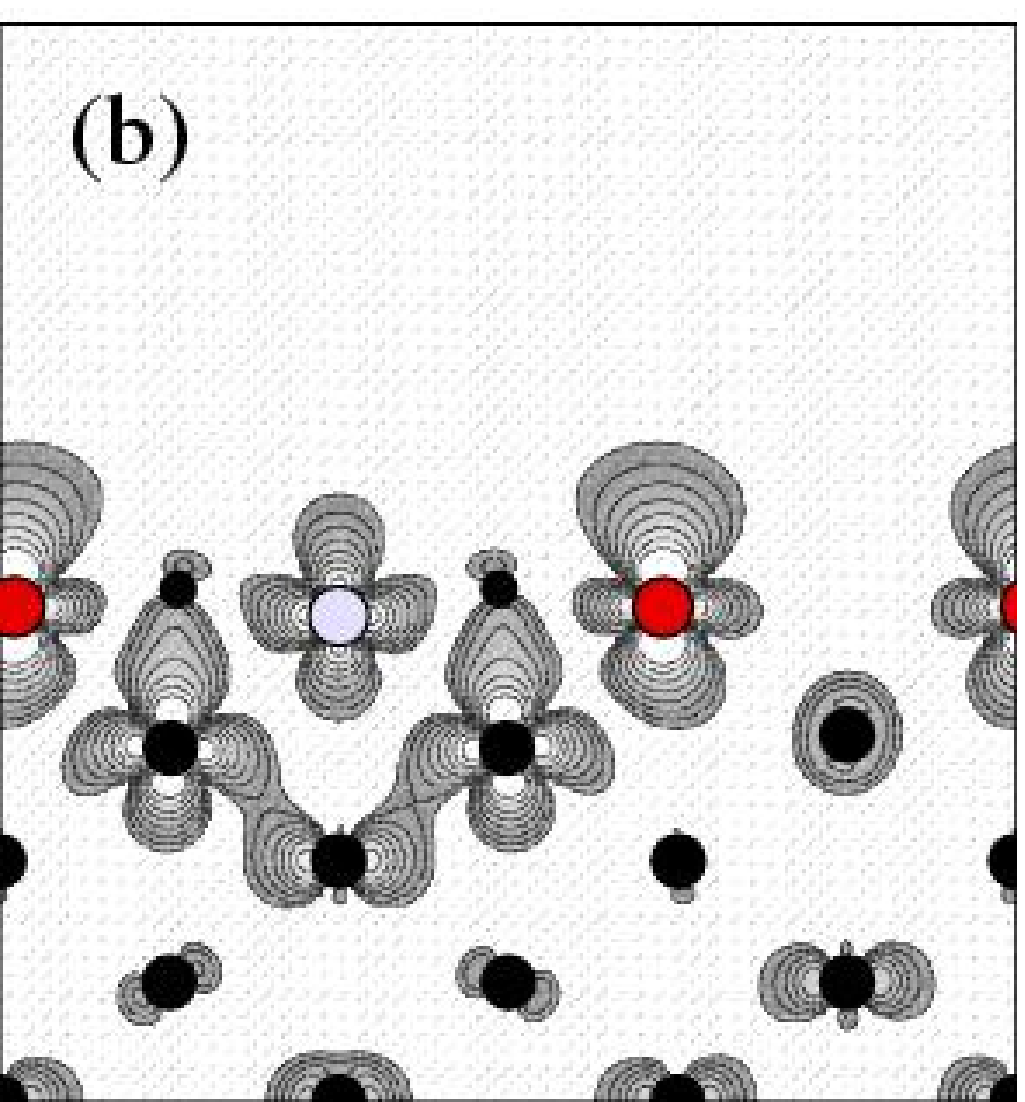, scale = 0.40}\hspace*{0.25cm}
    \epsfig{file=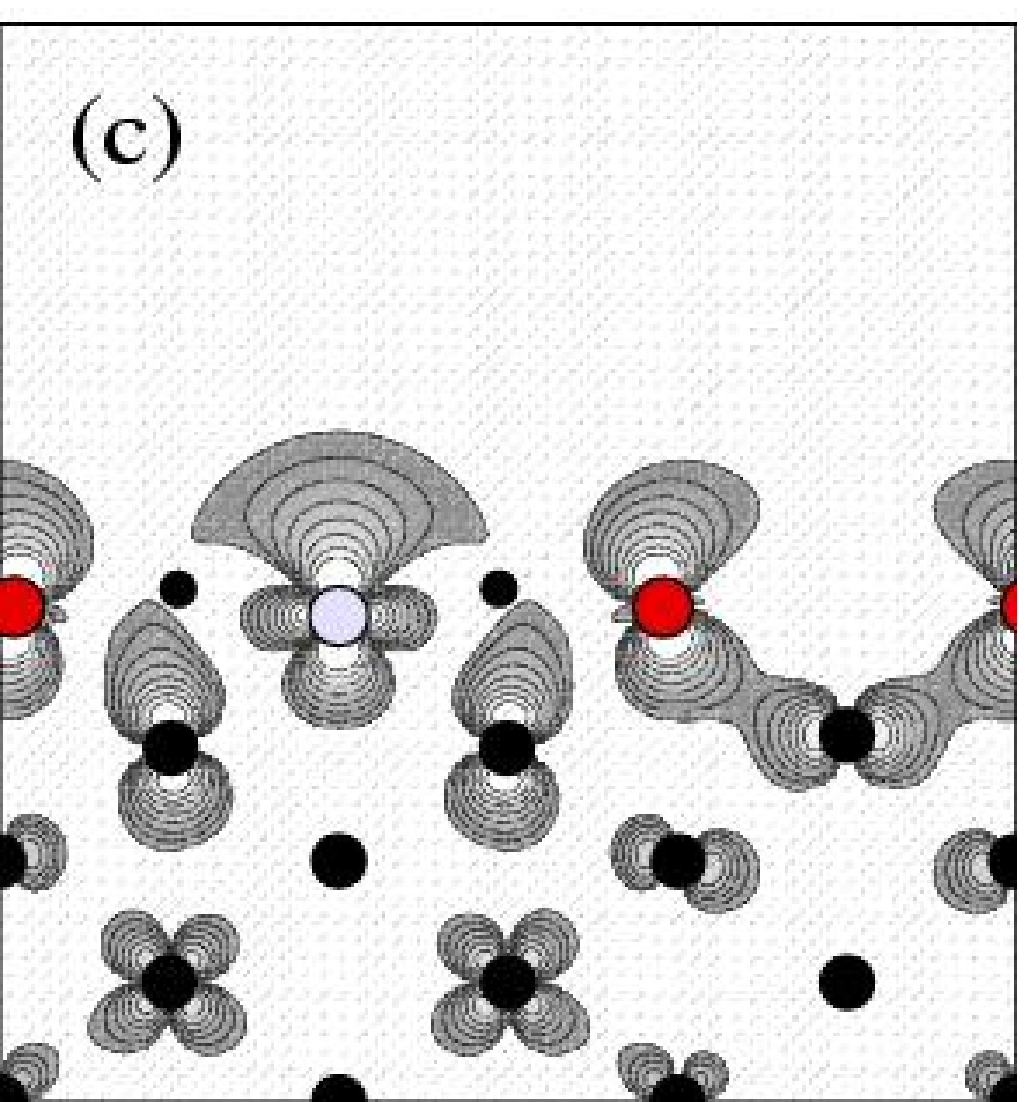, scale = 0.40}\hspace*{0.25cm}
    \epsfig{file=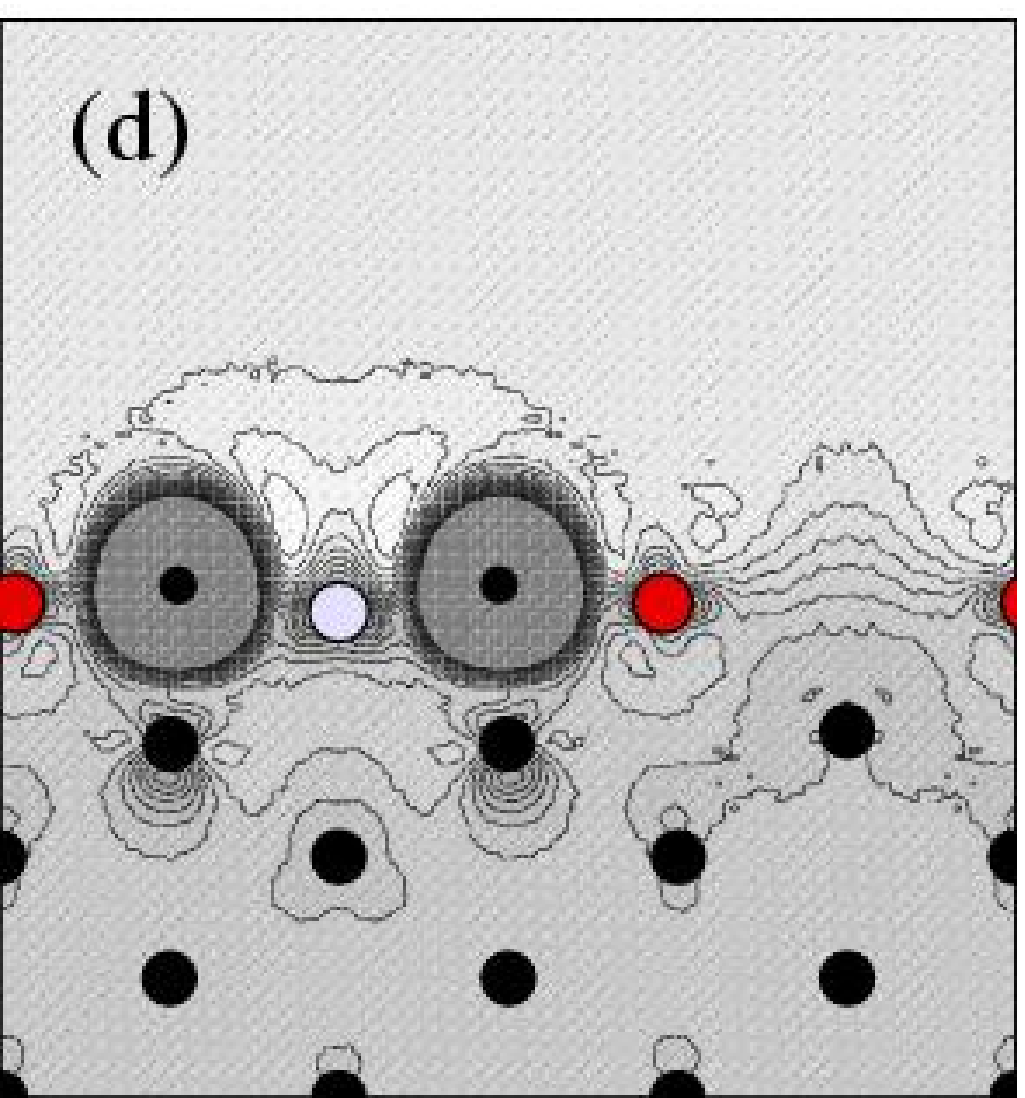, scale = 0.40}
    \caption[STATES]{\label{fig:states} (Color online) 
      Charge density of the minority spin states at 
      $\bar{\Gamma}$ with energy $0.43$ eV (a), $0.56$ eV (b) and $0.83$ eV (c). 
      The contour plots are displayed in an atomic plane perpendicular to both the 
      surface and the C chains. The C, Fe$_1$, and Fe$_2$ atoms are indicated 
      with small black disks, large red disks, and large light blue disks, respectively; 
      other iron atoms are indicated with large black disks. 
      A logarithmic scale common to the three plots has been used: the lowest contour 
      corresponds to $10^{-4}$ e/bohr$^{3}$ and the value increases by a factor of 
      $1.85$ from one countour to the next. 
      In panel (d) we display, in the same plane, the change in the electrostatic potential
      induced by the C atoms. The mean electrostatic potential is set to 
      zero in the bulk, and the contours are separated by 0.15 eV. The potential increases 
      from dark to light gray regions. Negative values are truncated below -1.8 eV.}
  \end{center}
\end{figure*}

In Fig.~\ref{fig:band}, we display the band dispersion of the 
minority-spin tunneling states with the largest integrated 
probability density in the vacuum beyond $d = 2.3$~\AA. 
The band dispersion is shown along the $\bar{\Gamma} - 
\bar{K}_{\|}^{BZ}$ line parallel to the C chains in the 
2-dimensional BZ of the $c(3 \sqrt{2} \times \sqrt{2})$ 
surface; $\bar{K}_{\|}^{BZ}$ stands for the zone-edge 
$k$ point $(\frac{\pi}{2a}, \frac{\pi}{2a})$. 
Above the Fermi energy, we find three bands of tunneling states 
which yield the main contribution to the LDOS features observed between 
0.3~eV and 1.0~eV in Fig.~\ref{fig:ldos}. 
The lowest-energy feature 
is associated with the two bands between 0.3~eV and 0.7~eV and 
the higher-energy feature with the band at $\sim 0.8$~eV. 
These three bands derive from minority-spin surface 
states of the clean Fe(001) surface, which persist at the 2/3 C 
covered surface. 
The probability densities of the corresponding three surface-band 
states at $\bar{\Gamma}$ are shown in Fig.~\ref{fig:states} (a), (b), 
and (c), in a $(1 \bar{1} 0)$ atomic plane perpendicular to the C 
chains. The two lowest-energy states include mainly $d_{3z^2-r^2}$ 
like orbitals of the Fe$_1$ atoms, whereas 
the higher-energy state involves predominantly  $d_{3z^2-r^2}$ like 
orbitals of the Fe$_2$ atoms. 
We note that these states are also mostly localized at the surface, as 
more than 60~\% of their electronic charge is located on the two 
outermost Fe atomic layers.

The minority-spin  $d_{3z^2-r^2}$ like surface states of the 
Fe(001) surface have been shown to induce a peak at about 
$0.2$~eV above $E_{\text{F}}$ in 
scanning tunneling spectroscopy (STS) of the clean Fe(001) 
surface.\cite{stroscio95,Biedermann96} In the presence of 
segregated C, we find that such states are shifted to higher energy, 
giving rise to the two features, between 0.3 and 1~eV, in the LDOS 
in Fig.~\ref{fig:ldos}. 
We note that similar shifts to higher energy of reminiscent surface 
states of the transition-metal substrate in the presence of segregated 
impurities have been observed at domain boundaries on the Si/Fe(001) 
surface alloy\cite{Biedermann96} and at contaminated C-rich areas on  
the V(001) surface.\cite{Bischoff01} 
In our case, however, we can distinguish two features: a 
low-energy feature, corresponding to minority-spin  Fe $d_{3z^2-r^2}$ like 
surface states localized within the C stripes, and a higher-energy 
feature, related mainly to Fe  $d_{3z^2-r^2}$ like states 
located in between the C stripes. 
The former states explain the 
zigzag chain structure observed in the simulated STM image (Fig.~\ref{fig:stm}c), 
while the latter are responsible for some additional weak features, located 
on the Fe$_2$ atoms, in the LDOS with increased $\Delta \varepsilon$ 
(Fig.~\ref{fig:stm}b).

Both the global shift to higher energy of the tunneling Fe 
surface states and their splitting into states located either within or in 
between the C stripes can be explained considering the changes in the   
electrostatic potential induced by the Carbon atoms at the surface. The 
energy shift is attributed to the change in the work function and the 
splitting to the induced corrugation of the electrostatic potential. 
The calculated work function of the Fe(001) surface changes from 4.0~eV, 
for the clean surface, to 4.8~eV, for the 2/3 C covered $c(3\sqrt{2} \times 
\sqrt{2})$  surface. This corresponds to a 0.8~eV increase in the 
electrostatic potential step at the surface in the presence of the carbon. 
This increase in the potential step occurs between $d \approx 0$ and $d 
\approx a$ (see Fig.~\ref{fig:states}d), and thus shifts surface 
states which extend far into the vacuum, and in particular the tunneling 
states of Fig.~\ref{fig:band}, to higher energy with respect to the 
Fermi energy. 
In Fig.~\ref{fig:states} (d), we display a contour plot, in the same 
$(1 \bar{1} 0)$ atomic plane as in Figs.~\ref{fig:states} (a,b,c), of the 
electrostatic potential change induced by the C atoms at the Fe 
$c(3\sqrt{2} \times \sqrt{2})$  surface. 
The corrugation induced by the presence of the C atoms gives 
rise to a region in the vacuum, in front of the Fe$_2$ stripes, 
in which the potential is globally repulsive with respect to the 
potential found, at the same distance from the surface, in front of the 
Fe$_1$ atomic rows. 
Such a corrugation of the electrostatic potential tends thus to shift 
the Fe $d_{3z^2-r^2}$ like surface states of the Fe$_2$ atoms to higher  
energy with respect to those of the Fe$_1$ atoms, leading to 
a localization of the resulting low/high energy surface states 
within/in between the C stripes. 

In summary, we predict one-dimensional minority-spin 
empty surface states near the Fermi energy at the C/Fe($001$) c($3\sqrt{2} \times 
\sqrt{2}$) surface. These states derive from $d_{3z^{2}-r^{2}}$-like surface 
states of the clean Fe(001) surface and are laterally confined within the  
self-assembled carbon chains. 
Simulation of the STM images indicates that the zigzag structure observed in 
STM experiments at positive biases corresponds to the 
zigzag chains of iron atoms enclosed within the C stripes, and that the 
STM signal is largely due to the one-dimensional Fe surface states.
Furthermore, we predict that the one-dimensional minority-spin surface states 
gives rise to a significant feature at $\sim0.6$~eV above the Fermi energy, which 
should be observable in (spin-resolved) STS experiments.
\\

We thank J. Fujii, G. Panaccione, I. Vobornik and G. Rossi for having drawn
our attention on this system and for many helpful discussions. We would 
like also to thank M. Altarelli, P. Messina, and N. Stojic for very useful 
discussions. One of the authors (G. T.) acknowledges the support by Agatino 
Trimarchi during the first phase of this project. The calculations
in this work have been carried out using the PWscf package.\cite{pwscf} 
\end{document}